\documentclass[preprint,showpacs,preprintnumbers,amsmath,amssymb,prl]{revtex4}

\usepackage{graphicx}
\usepackage{dcolumn}
\usepackage{bm}

\begin{document}

\title{Magnetic field dependence of the quantum tunneling of normal-superconductor interfaces in a type-I Pb superconductor}

\author{Sa\"{u}l V\'{e}lez}
\author{Ricardo Zarzuela}
\author{Antoni Garc\'{i}a-Santiago}
\email{agarciasan@ub.edu}
\author{Javier Tejada}

\affiliation{Grup de Magnetisme, Departament de F\'{i}sica Fonamental, Facultat de F\'{i}sica, Universitat de
Barcelona, c. Mart\'{i} i Franqu\`{e}s 1, planta 4, edifici nou, 08028 Barcelona, Spain
\\
and Institut de Nanoci\`{e}ncia i Nanotecnologia IN$^2$UB, Universitat de Barcelona, c. Mart\'{i} i Franqu\`{e}s 1,
edifici nou, 08028 Barcelona, Spain}

\date{\today}

\begin{abstract}
We report experimental evidence of the effect of an applied magnetic field on the non-thermal magnetic relaxation in a disk-shaped type-I lead superconductor. The time evolution of the irreversible magnetization proves to be logarithmic for a wide range of temperatures and magnetic field values along the descending branch of the hysteresis cycle. When the intensity of the magnetic field increases, the crossover temperature separating the thermal and non-thermal regimes of magnetic relaxation is found to decrease, whereas the rate at which such relaxation occurs is observed to increase. These results are discussed in the framework of a recent model for quantum tunneling of normal-superconductor interfaces through the distribution of pinning energy barriers generated by structural defects in the sample, considering that the strength of the barriers decreases with the magnetic field. A phase diagram describing the dynamics of interfaces during flux expulsion in the intermediate state as a function of temperature and magnetic field is constructed.
\end{abstract}


\maketitle

\section{Introduction}

The properties of type-I superconductors have been studied, both theoretically and experimentally, for nearly one century \cite{Landau,Liv,Gold1,Gold2,Cebers,Gourd,Proz2005,ProzET,Suprafroth,Proz2008,Menghini,Vel1,Rotating,QTI,Vel2}.
Historically, the presence of irreversible phenomena in the magnetic hysteresis cycle had been associated to physical or chemical defects such as dislocations or impurities \cite{Liv}. However, recent works by Prozorov and coauthors \cite{Proz2005,ProzET,Suprafroth}, using high-resolution magneto-optical imaging and magnetic measurements in extremely pure, defect-free type-I lead (Pb) superconductors in the intermediate state, have unveiled that the dynamics of the system can exhibit an intrinsic irreversibility correlated with the existence of different topologies for the flux structures. This striking feature can be only observed when the external magnetic field is parallel to the revolution axis of the sample, whereas the magnetization of the system becomes fully reversible if the magnetic field is perpendicular to this axis \cite{Vel1}. The existence of these phenomena has been attributed to the presence of a geometrical barrier, which controls both the penetration and the expulsion of the magnetic flux. To be precise, when the magnetic field increases from zero the flux penetrates in the form of bubbles that tend quickly to fill the whole sample to reach the so-called suprafroth state \cite{Suprafroth}, but when the magnetic field is reduced from the normal state the system exhibits random labyrinthine patterns that make the expulsion to be more difficult than the penetration \cite{ProzET,Vel1}.

The effect of the geometrical barrier is fully suppressed at zero magnetic field, indicating that the system is not able to retain any flux line at this intensity \cite{ProzET}. However, if the sample also contains other irreversibility sources, metastability develops and a good amount of magnetic flux can be trapped at zero field \cite{Vel1,Vel2}. The isothermal evolution of such remnant state has been reported to follow a logarithmic law with time in flat disks of Pb with stress defects \cite{QTI}. Evidence of a mechanism of quantum nature as responsible for the relaxation of the system was found below a certain crossover temperature. Resemblances between the movement of normal-superconductor interfaces (NSI) in type-I superconductors and that of domain walls in ferromagnets suggest that the logarithmic relaxation observed so far in Pb samples is brought about by the dissipation of the NSI traversing a broad distribution of local pinning potentials associated with the defects of the sample \cite{QTI}. A model for quantum tunneling of interfaces (QTI) based on the Caldeira-Leggett theory for quantum dissipation \cite{Legget} was recently developed considering that the magnetization evolves with time due to the formation and decay of mesoscopic bumps in the NSI, and was found to fit rather well the experimental data \cite{QTI}. The pinning barriers act as a source for metastability along the descending branch of the hysteresis cycle and are responsible for the temperature-dependent irreversibility exhibited by the samples in which both geometrical barrier and stress defects are present \cite{Vel2}. The aim of this paper is to investigate the effect of an external magnetic field on the dynamics of NSI in a type-I Pb superconductor. The crossover between thermal and non-thermal regimes will be probed and discussed in terms of the QTI model.

\section{Experimental Set-up}

The sample investigated was a thin disk of extremely pure (99.999 at.$\%$) type-I superconducting Pb with a surface area of 40 mm$^2$ and a thickness of 0.2 mm, whose preparation and magnetic characterization have been given elsewhere \cite{QTI}. Magnetic measurements were carried out in a commercial superconducting quantum interference device (SQUID) magnetometer at temperature values from $T =$ 1.80 K to $T =$ 7.00 K with a low temperature stability better than 0.01 K \cite{SQUID}, by applying the magnetic field, $H$, always parallel to the revolution axis of the sample, with intensities up to 1 kOe. Isothermal magnetization curves, $M(H)$, were measured after the sample had been first zero-field-cooled (ZFC) from the normal state down to the desired temperature. At each $T$ value, the first magnetization curve, $M_{\rm{1st}}(H)$, was measured by sweeping $H$ from zero up to the normal state, and the descending branch of the corresponding hysteresis cycle, $M_{\rm{des}}(H)$, was measured by subsequently sweeping $H$ back to zero. The magnetic field dependence of the field-cooled (FC) magnetization, $M_{\rm{FC}}(H)$, was also registered at different temperatures by cooling the sample from the normal state down to each $T$ value under different values of $H$. Finally, the isothermal time evolution of the remnant magnetization obtained at different $H$ values along $M_{\rm{des}}(H)$ was recorded. The results were analyzed in terms of the reduced magnitudes $h \equiv H/H_c$ and $m \equiv M/H_c$, where $H_c(T) = H_{c0}[1 - (T/T_c)^2]$ is the thermodynamic critical field and $T_c$ is the superconducting transition temperature \cite{Rose-Innes}. Fitting the expression of $H_c(T)$ to experimental data obtained from the $M(H)$ curves produced values of $H_{c0} =$ 802 $\pm$ 2 Oe and $T_c =$ 7.23 $\pm$ 0.02 K for our sample, which agree fairly well with typical values found in similar samples \cite{landolt}.

\section{Results}

Fig. 1 shows a detail of the $m_{\rm{1st}}(h)$ curves and $m_{\rm{des}}(h)$ branches measured at $T=$ 2.00 K (solid squares) and $T=$ 6.00 K (solid circles), in the low-$h$ regime ($h_{\rm{max}} \simeq$ 0.35 in the figure), where the magnetic hysteresis cycles exhibit the strongest irreversibility. In this region, the two $m_{\rm{1st}}(h)$ curves largely superimpose whereas the $m_{\rm{des}}(h)$ branches depend on the value of $T$. Actually, the $m_{\rm{des}}(h)$ branches obtained as $T$ increases progressively from 2.00 K to 6.00 K (not shown here) span the space bound by the two branches in the figure, while all the corresponding $m_{\rm{1st}}(h)$ curves scale onto a single one, in good agreement with previous results obtained in similar samples \cite{Vel2}. Fig. 1 also presents the $m_{\rm{FC}}(h)$ values obtained at $T=$ 2.00 K (open squares) and $T=$ 6.00 K (open circles). Both curves practically superimpose in this representation in a similar way as the two $m_{\rm{1st}}(h)$ curves do.

\begin{figure}[htbp!]
\includegraphics[scale=0.6]{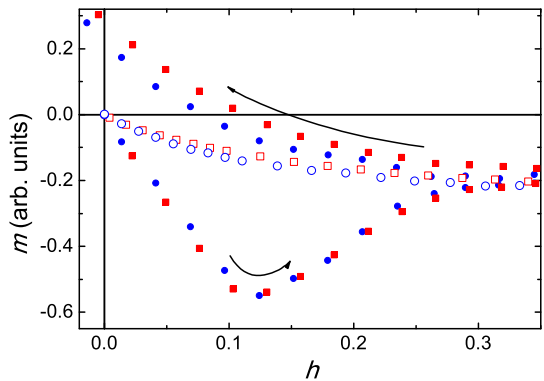}
\caption{(Color online) First magnetization curves (left-to-right arrow) and descending branches (right-to-left arrow) of the hysteresis cycles measured at 2.00 K (solid squares) and 6.00 K (solid circles) when the magnetic field is applied parallel to the revolution axis of the sample after a ZFC process. The magnetic field dependence of the FC magnetization at the same temperatures is included (open squares for 2.00 K, open circles for 6.00 K). The data are plotted using the reduced $m(h)$ representation defined in the text. The region of $h$ values up to 0.35 has been chosen to show the irreversible part of the hysteresis cycles in detail.}
\label{fig1}
\end{figure}

\begin{figure}[htbp!]
\includegraphics[scale=0.6]{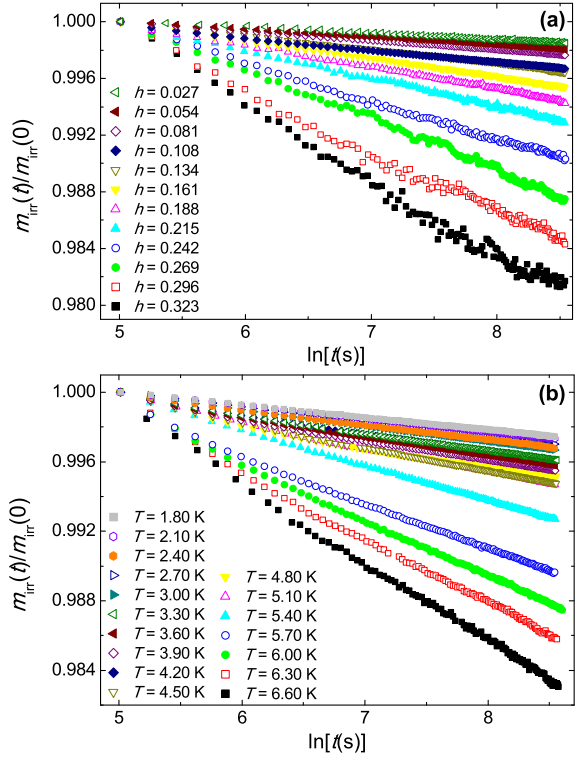}
\caption{(Color online) Logarithmic time evolution of the normalized reduced irreversible magnetization of the sample, $m_{\rm{irr}}(t)/m_{\rm{irr}}(0)$, measured for different $h$ values from 0.027 (uppermost curve) to 0.323 (lowermost curve) in steps of 0.027 at $T=$ 2.00 K [panel (a)], and at different $T$ values from 1.80 K (uppermost curve) to 6.60 K (lowermost curve) in steps of 0.30 K for $h=$ 0.10 [panel (b)].}
\label{fig2}
\end{figure}

In a defect-free sample, in which only geometrical effects are the source for irreversibility, a large energy barrier separates the states that can be attained by submitting the sample to different magnetic history processes and provides strong stability to the system \cite{ProzET}. At a certain temperature, $m_{\rm{1st}}(h)$ should thus follow a series of equilibrium states corresponding to the flux penetration as $h$ increases, while $m_{\rm{des}}(h)$ and $m_{\rm{FC}}(h)$ should actually coincide because both curves would correspond to a succession of equilibrium states that would be followed during the flux expulsion as the magnetic field decreases \cite{Proz2005}. The lack of isothermal relaxation along the whole $m(h)$ curve and the thermal independent behavior of $m_{\rm{des}}(h)$ \cite{Vel2} in a sample of this kind are strong proofs of this fact. When other sources of irreversibility like stress defects are present, the ability of the system to trap magnetic flux during the expulsion is enhanced with respect to the defect-free case and is substantially influenced by temperature \cite{Vel2}. This corroborates the differences observed in Fig. 1 among the $m_{\rm{des}}(h)$ branches at different $T$ values and between these and $m_{\rm{FC}}(h)$. In this case, only the latter would follow the energy minima for flux expulsion, whereas $m_{\rm{1st}}(h)$ would still define the equilibrium for flux penetration, as it is confirmed by the fact that the magnetization did not evolve with time starting anywhere along this curve \cite{Vel2}.

In this context, the pinning energy barriers associated with defects should provide a multiplicity of metastable states along $m_{\rm{des}}(h)$ that would originate time-dependent phenomena just as in magnetic materials \cite{BookMQTMM}. For a broad distribution of barriers, as it should be expected in our sample according to the preparation procedure, the reduced irreversible magnetization, $m_{\rm{irr}}$, should evolve with time following a logarithmic dependence \cite{QTI}. Such magnetization was defined as $m_{\rm{irr}}(t) \equiv m(t)-m_{\rm{eq}}$ and was considered as the amount of reduced magnetization, $m(t)$, that deviates from the magnetic equilibrium state, $m_{\rm{eq}}$, which is given by the appropriate $m_{\rm{FC}}(h)$ value. Panel (a) in Fig. 2 shows $m_{\rm{irr}}(t)/m_{\rm{irr}}(0)$ as a function of $\ln(t)$ for several $h$ values from 0.027 (uppermost curve) to 0.323 (lowermost curve) in steps of 0.027 at $T=$ 2.00 K, while panel (b) shows the same representation at different $T$ values from 1.80 K (uppermost curve) to 6.60 K (lowermost curve) in steps of 0.30 K for $h=$ 0.10. A good linear dependence is observed for all curves in both panels and was also found for all combinations of $h$ and $T$ values explored. The whole data set was fitted to the law $m_{\rm{irr}}(t) = m_{\rm{irr}}(0)[1-S \ln(t)]$, to determine the dependence of the magnetic relaxation rate on temperature and reduced magnetic field, $S(T,h)$.

\begin{figure}[htbp!]
\includegraphics[scale=0.6]{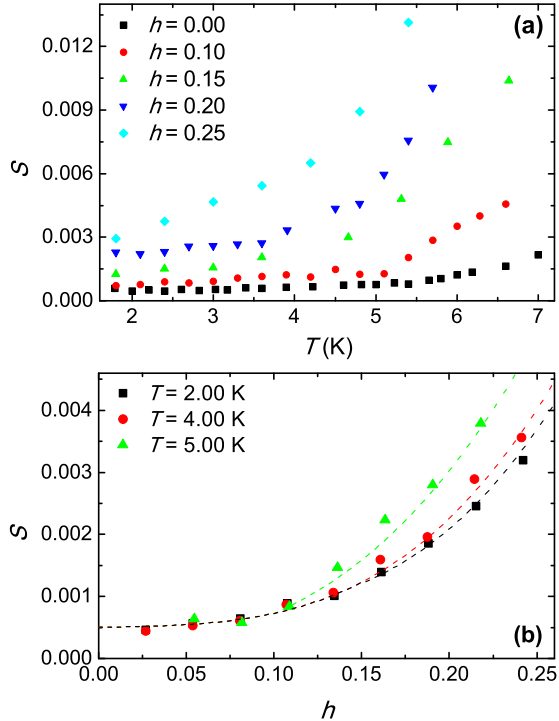}
\caption{(Color online) (a) Temperature dependence of the magnetic relaxation rate, $S(T)$, obtained for $h=$ 0.00 (squares), $h=$ 0.10 (circles), $h=$ 0.15 (upward triangles), $h=$ 0.20 (downward triangles), and $h=$ 0.25 (rhombuses). (b) Reduced magnetic field dependence of the magnetic relaxation rate, $S(h)$, obtained for $T=$ 2.00 K (squares), $T=$ 4.00 K (circles), and $T=$ 5.00 K (upward triangles). The dashed lines are guides to the eye.}
\label{fig3}
\end{figure}

Panel (a) in Fig. 3 shows the $S(T)$ curves obtained for $h=$ 0.00 (squares), $h=$ 0.10 (circles), $h=$ 0.15 (upward triangles), $h=$ 0.20 (downward triangles), and $h=$ 0.25 (rhombuses), while panel (b) shows the $S(h)$ curves obtained at $T=$ 2.00 K (squares), $T=$ 4.00 K (circles), and $T=$ 5.00 K (upward triangles). As $T$ increases, two regimes can be identified in panel (a) for all $h$ values but 0.25: $S$ exhibits a practically constant plateau, $S_Q$, up to some critical temperature $T_Q$, above which it increases with $T$. The behavior of $S(T)$ in the second regime is the common signature of thermal activation processes, while the occurrence of $S_Q$ is an indication that magnetic relaxation in the first regime should be ascribed to quantum phenomena instead \cite{BookMQTMM}. The value of the plateau changes progressively from $S_Q$(0.00) $\sim$ 0.0005 to $S_Q$(0.20) $\sim$ 0.0020, while the crossover temperature tends to decrease from $T_Q$(0.00) $\sim$ 5.4 K to $T_Q$(0.20) $\sim $ 4.0 K. At the same time, the slope of $S(T)$ above $T_Q$ grows as $h$ increases. In particular, only the increasing regime in $S(T)$ can be observed for $h =$ 0.25 in the whole temperature range of our experiments. The shape of the curve in this case prompts to estimate $T_Q$ at somewhere around 2.0 K. On the other hand, two successive regimes can be also distinguished in panel (b): as $h$ increases, the three curves superimpose onto a single one that rises slowly up to $h \sim$ 0.12, while above this value they separate progressively and become steeper. In particular, the curve at $T=$ 5.00 K deviates at $h \sim$ 0.12, while the curve at $T =$ 4.00 K deflects at $h \sim$ 0.17, indicating that these $h$ values determine the crossover reduced magnetic field, $h_Q$, that separates the quantum and thermal regimes at these temperatures. In conformity with what has been remarked for panel (a), the first increasing region of $S(h)$ in panel (b) shows actually the dependence of the tunneling rate on the reduced magnetic field, $S_Q(h)$.

\section{Discussion}

The origin of a remnant magnetic state that relaxes towards equilibrium lies in the capability of stress defects to pin the NSI when the magnetic field is reduced from the normal state in a magnetic history dependent process. The onset of such state along $m_{\rm{des}}(h)$ takes place at a certain temperature-dependent magnetic field intensity, the so-called reduced irreversibility field, $h^*(T)$, that can be identified experimentally as the point at which $m_{\rm{des}}(h)$ departs from $m_{\rm{1st}}(h)$ \cite{Vel2}. As $h$ decreases at constant temperature, $h^*$ should separate a regime that would be characterized by quasi-free flux motion (at $h \gtrsim h^*$) from another regime in which the time evolution of the magnetization would occur via thermal activation of NSI over the pinning energy barriers (at $h \lesssim h^*$). On the other hand, as it has been discussed above, $T_Q(h)$ describes a magnetic-field-dependent crossover temperature that separates the thermal and quantum regimes of magnetic relaxation. This can be also understood as a temperature-dependent transition value for the magnetic field, $h_Q(T)$, as it has been established in Fig. 3(b). According to this, we may construct the phase diagram that describes the dynamics of the NSI along the $m_{\rm{des}}(h)$ curve, as it is shown in Fig. 4. The two lines in the diagram [squares for $h^*(T)$, circles for $T_Q(h)$ and $h_Q(T)$] show decreasing tendencies that indicate that both the thermal and quantum regimes of depinning of the NSI move to lower temperatures as $h$ increases. This actually implies that the increment of the intensity of the applied magnetic field has an important influence on the reduction of the strength of the pinning energy barriers that control metastability \cite{Vel2}. The motion of NSI along the sample would be therefore less obstructed by defects and, as a consequence, the magnetization should evolve faster with time. This would explain why the slope of $S(T)$ in the thermal regime and the value of $S_Q$ grow as $h$ increases in Fig. 3.

\begin{figure}[htbp!]
\includegraphics[scale=0.6]{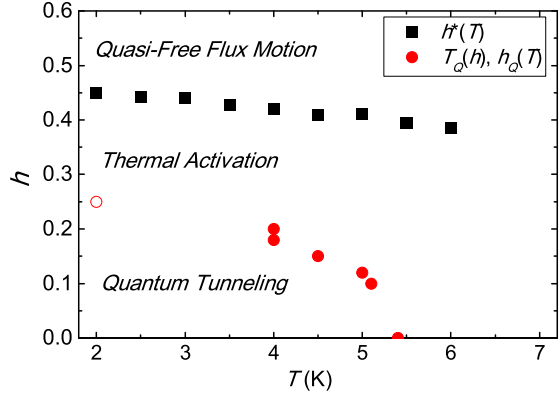}
\caption{(Color online) Reduced magnetic field vs. temperature phase diagram showing the different regimes that describe the dynamics of the normal-superconductor interfaces when the initial state lies along the descending branch of the hysteresis cycle, $m_{\rm{des}}(h)$. Solid squares correspond to the irreversibility line, $h^*(T)$, that separates the regimes of quasi-free flux motion and thermal activation, while solid circles correspond to the crossover line, $T_Q(h)$ and $h_Q(T)$, that separates the thermal and quantum regimes of magnetic relaxation. The open circle stands for the value of $T_Q$ estimated for $h =$ 0.25.}
\label{fig4}
\end{figure}

We will now try to understand our results in the framework of the QTI model. Magnetic relaxation in type-I superconductors occurs due to the formation and decay of bumps at the NSI when they are pinned at the defects. These bumps are characterized by a lateral size $L$ and a height $a$ (see Fig. 6 in Ref. \onlinecite{QTI}), which should both depend on $h$. The energy barrier associated with the bump is independent of $L(h)$ and can be generally expressed as $U_B(h)=\sigma\pi [a(h)]^2$, where $\sigma = \xi B_c^2 / (3\sqrt2 \pi)$ is the surface energy density of the NSI, which actually does not depend on the magnetic field, $B_c$ is the critical magnetic induction, and $\xi$ is the superconducting coherence length. The theoretical expression for the relaxation rate in the quantum regime is given by $S_{Q}(h)=A(h)e^{-I_{eff}(h)/\hbar}$, where the prefactor can be assumed to contribute much less than the exponent, which is dominated by the dissipative term of the Caldeira-Leggett effective action, $\displaystyle I_{eff}(h)\approx(\eta/4\pi)[L(h)a(h)]^2$ \cite{QTI}. Here $\eta = \sqrt{\lambda_L\xi}B_{c}^2/(2\rho_{n}c^2)$ is the drag coefficient, $\lambda_L$ is the London penetration depth, and $\rho_{n}$ is the normal-state resistivity, which can be considered magnetic field independent in our experimental conditions \cite{Berg}. Consequently, $\eta$ does not depend on $h$, and $L(h)$ and $a(h)$ can be established in terms of their zero magnetic field values as

\begin{equation}
\label{eq1}
\frac{U_{B}(h)}{U_{B}(0)}=\left[\frac{a(h)}{a(0)}\right]^2, \quad\frac{T_{Q}(h)}{T_{Q}(0)}=\left[\frac{L(0)}{L(h)}\right]^2,
\end{equation}
where the energy barrier and the crossover temperature are related via the expression $T_{Q}=\hbar U_{B}/I_{eff}$. As it has been discussed, both $U_B$ and $T_Q$ decrease as $h$ increases, and this translates through Eq. \eqref{eq1} into a reduction of the height $a(h)$ and an enlargement of the lateral size $L(h)$ of the bump. Therefore, the main effect of the increment of the magnetic field is to flatten the bumps of the NSI pinned by the defects, so for large values of $h$ the energy barrier becomes weak enough to avoid the experimental observation of $T_Q$, as it happens in Fig. 3(a) for $h =$ 0.25.

Finally, the increasing dependence observed experimentally for $S_Q(h)$ in Fig. 3 implies that $I_{eff}(h)$ should be a decreasing function. Considering the expression given above for $I_{eff}(h)$ and taking into account Eq. \eqref{eq1}, we may then infer that the ratio $U_B(h)/T_Q(h)$ should also show a decreasing dependence. As both $T_Q$ and $U_B$ diminish with $h$, if this relation is to be fulfilled in the $h$ range of Fig. 3, the variation of $U_B(h)$ should be steeper than that of $T_Q(h)$. This appears to be supported by the behavior of the curves in Fig. 3(a): as $h$ goes from 0.00 to 0.20, the decrease in the value of $T_Q$ turns out to be slower than the increase in the slope of $S(T)$ above $T_Q$, which should be in fact related with the decrease of $U_B$ if thermal activation processes are positively involved in this regime. Further experimental work is needed to confirm the validity of this conjecture.

\section{Conclusions}

To summarize, the nature of the mechanism of magnetic relaxation in a disk-shaped type-I Pb superconductor has been explored as a function of temperature and magnetic field for initial states along the descending branch of the magnetic hysteresis cycle. As the sample is cooled down, a transition from a thermal to a quantum regime has been observed in the temperature dependence of the magnetic relaxation rate. The crossover temperature between both regimes has been found to exhibit an inverse dependence on the magnetic field that impedes the detection of the quantum regime at high enough magnetic field values. On rising the magnetic field, the relaxation rate has been observed to change from a slowly increasing regime, in which no thermal effects are detected, to a steeply incremental region that is noticeably affected by temperature. Comparison between the experimental data and a model for quantum tunneling of normal-superconductor interfaces through energy barriers defined by structural defects in the sample implies that the bumps generated at the interfaces, which are responsible for the magnetic relaxation, become flatter when the magnetic field increases. Quantum transitions cease above a certain combination of values of temperature and magnetic field, but the interfaces are still pinned by the defects and depinning occurs by thermal activation until a threshold for quasi-free flux motion is reached. Plotting the values of reduced magnetic field and temperature that characterize these transitions, a phase diagram with the different dynamical regimes describing the motion of interfaces during flux expulsion in the intermediate state has been obtained.

\section{Acknowledgments}

S. V. and R. Z. acknowledge financial support from Ministerio de Ciencia e Innovaci\'{o}n de Espa\~{n}a. A. G.-S. thanks Universitat de Barcelona for backing his research. J. T. appreciates financial support from ICREA Academia. This work was funded by the Spanish Government project MAT2008-04535.

\bibliography{Type-I}

\end{document}